\documentstyle[12pt,twoside,fleqn]{article}    

\title{Quantum generalization of the classical rotating solutions of 
the O(N) model}
\author{D.~Vautherin$^1$ and T. Matsui$^2$ \\ \\
$^1$ Physique Th\'eorique des Particules 
El\'ementaires \\ Universit\'e Pierre et Marie Curie \\ 
F-75252, Paris Cedex 05, France \\ \\
$^2$ Yukawa Institute for Theoretical Physics \\ 
Kyoto University, Kyoto 606, Japan}
\date{\small May, 1998; Preprint YITP-98-28}
\begin{document}
\maketitle

\begin{abstract}
Analytic solutions of the mean field evolution equations for an
N-component scalar field with O(N) symmetry are presented. These
solutions correspond to rotations in isospin space. They represent
generalizations of the classical solutions obtained earlier by Anselm
and Ryskin.  As compared to classical solutions new effects arise
because of the coupling between the average value of the field and
quantum fluctuations.
\end{abstract}

\vfill\eject

\section{Introduction}
\label{sect_introduction}

A class of analytic solutions to the classical equations of motion of
the non-linear sigma model have been constructed by Anselm and
Ryskin\cite{ANSELM} who conjectured the possible relevance of such
solutions to high energy heavy ion collisions where a large number of
pions are produced in the final states and a part of them may be
described by a coherent state wave function.  Subsequently, Blaizot
and Krzywicki\cite{BLAIZOT} presented another type of classical
time-dependent solutions which possess a symmetry under the
Lorentz-boost in the direction of colliding nuclei and used them to
describe the soft pion emission in ultra-relativistic nuclear
collisions.  Interest in these solutions grew up because of their
possible relevance for the observation of a disoriented chiral
condensate\cite{BJORKEN,RAJAGOPAL,DEVEGA} which may be formed when the
chiral symmetry, temporarily restored in hot hadronic matter produced
by the nuclear collision, gets broken again by the rapid cooling of
the matter by expansion.

These classical oscillating solutions correspond, in the quantum
language, to coherent states of quantum fields, rotating in the
internal symmetry space of the fields.  One may consider them as
arising due to the spontaneous breaking of the internal symmetry,
similar to nuclear deformation and rotation.  In the latter case, it
is known that the nuclear rotation affects the nuclear deformation
through the coupling of the collective motion of the deformed nuclear
mean field to the motions of intrinsic nucleonic degrees of freedom in
it\cite{THOULESS}.
It is the purpose of this note to show that similar phenomenon arises
in the case of oscillating coherent quantum fields when their dynamics
is treated in the mean field approximation.  For this purpose we use
the framework of the time-dependent mean field theory of the quantum
fields which can be derived from the Gaussian Ansatz to the
time-dependent variational wave functional of the functional
Schr\"odinger equation\cite{JACKIW,VAUTHERIN}:
\begin{eqnarray*} 
\Psi \left[ \varphi ; t \right] 
& \equiv & \langle \varphi | \Psi (t) \rangle \\ & = & {\cal
N} \exp \left( i \langle \bar \pi | \varphi - \bar \varphi \rangle \right)
\\ && \times \exp \left( - \langle \varphi- \bar \varphi | ( \frac{1}{4G} + i
\Sigma ) | 
\varphi- \bar \varphi \rangle \right) \mbox{,}
\end{eqnarray*}
where $\varphi ({\bf x}) = \langle {\bf x} | \varphi \rangle $ is the
coordinate of the quantum field and the $G$, $\Sigma$, $\bar \varphi$,
$\bar \pi$ define respectively the (time-dependent) real and imaginary
part of the kernel of the Gaussian and its average position and
momentum.  Although this approach appears to sacrifice the covariance
with respect to Lorentz transformation, one can show that the
resultant equations of motion can be rewritten in a covariant form and
becomes equivalent to the ones obtained from an explicitly covariant
formulation\cite{CORNWALL}.

\section{Mean Field Equations}
\label{sect_mf}

For generality we consider an N-component scalar field $\varphi_a (
{\bf x})$, $a$=1,\ldots N with O(N) symmetry, characterized by a bare
mass $m_0^2$ and a coupling constant $\lambda$.  The index $a$ will be
referred to in what follows as flavor or isospin index. In the mean
field approximation the evolution of the mean value ${\bar \varphi}$
of the field (often called condensate) \cite{VAUTHERIN}
\begin{displaymath}
{\bar \varphi}_a(x) = \langle \Psi(t) | \varphi_a({\bf x})| 
\Psi(t) \rangle ,
\end{displaymath} 
is governed by the following equation of
motion
\begin{equation} \label{MEANFIELD}
\left[ \Box+ m_0^2 + \frac{\lambda}{6} {\bar \varphi}^2 +
\frac{\lambda}{6} {\rm tr} S(x,x)+
\frac{\lambda}{3} S(x,x) \right] {\bar \varphi} (x) =0  \mbox{.}
\end{equation} 
where $x$= $(t, {\bf x} )$, ${\bar \varphi} (x)$ stands for the
N-component vector ${\bar \varphi}_a (x)$, and ${\bar \varphi}^2 =
{\bar \varphi}_1^2$ + \ldots + ${\bar \varphi}_N^2$.  Here it is
implicit that the first four terms carry the N$\times$N unit
matrix. In the previous equation, $S$ is a N$\times$N matrix operator
in flavor space which is related to the kernel of the Gaussian
wave-functional
\begin{displaymath}
S (x, x) = \langle {\bf x} | G (t) | {\bf x} \rangle 
\end{displaymath} 
and the trace runs over the flavor indices.  $S (x, y)$ is the
Feynman propagator which satisfies 
\begin{displaymath}
\left[ \Box_x + m^2(x) \right] S( x, y ) = i \delta^4 (x-y), 
\end{displaymath}
with the boundary condition $S (x, y) = S (y, x)$; it is symbolically
written by
\begin{equation}\label{PROPAGATOR}
S = \frac{i}{ \Box + m^2(x) - i\epsilon }
\end{equation}
where the N$\times$N mass matrix is  
\begin{eqnarray} \label{MASS}
m^2 (x) & = & m_0^2 +\frac{\lambda}{6} {\bar \varphi}^2(x) + 
\frac{\lambda}{6} {\rm tr} S(x,x) \nonumber \\
& & \qquad + \frac{\lambda}{3} {\bar \varphi}(x) \times {\bar
\varphi}(x) + \frac{\lambda}{3} S(x,x) \mbox{.}
\end{eqnarray}
In this expression the symbol ${\bar \varphi} (x) \times {\bar
\varphi} (x)$ denotes the N$\times$N matrix whose $(a,b)$ matrix
element is ${\bar \varphi}_a(x) {\bar \varphi}_b(x)$.  The previous
equations (\ref{MEANFIELD}) -- (\ref{MASS}) are
non-linear because the motion of the condensate involves the mass
matrix $m^2(x)$ which in turn involves the values of the condensate and
of the propagator. A partial differential form of the
mean field equations can be written down by introducing the so-called
mode functions\cite{HOLMAN}. Note that the first three terms in
equation (\ref{MEANFIELD}) correspond to the classical approximation
considered by Anselm and Ryskin.  The next two correspond to the
contribution of quantum fluctuations whose effect is the object of the
present study.

\section{Rotating Solutions}
\label{sect_rs}

Let us look for solutions of the previous equations by means of the
following Ansatz
\begin{displaymath}
{\bar \varphi}(x)= U(x) {\bar \varphi}^{(0)}=
\exp \{i ( q \cdot x ) \tau_y \} {\bar \varphi}^{(0)},
\end{displaymath}
where $q_{\mu} =(\omega, {\bf q})$ and $\tau_y $ is a generator of
rotation in the subspace of flavor 1 and 2:
\begin{displaymath}
\tau_y = \left( 
\begin{array}{cccc}
0 & -i & 0 & \cdots \\
i & 0  & 0 & \cdots \\
0 & 0  & 0 & \cdots \\
\vdots & \vdots & \vdots & \ddots 
\end{array}
\right) 
\end{displaymath}
${\bar \varphi}^{(0)}$ can be interpreted as the mean field in the
{\em rotating frame} and we assume it to be independent of space and
time, pointing in the direction of the first flavor:
\begin{equation}\label{PHI0}
{\bar \varphi}^{(0)} = \left(
\begin{array}{c} 
\varphi_0 \\
0 \\
\vdots \\ 
\end{array}
\right)
\end{equation}
We also introduce the propagator $S^{(0)}$ in the rotating frame by
\begin{displaymath}
S(x,y)= U(x) S^{(0)} (x,y) U^{\dagger}(y),
\end{displaymath} 
with
\begin{displaymath} 
S^{(0)} = \frac{i}
{(\partial_{\mu} + i q_{\mu} \tau_y)(\partial^{\mu} + i q^{\mu} \tau_y)
+ M^2 - i \varepsilon}.
\end{displaymath}
We assume that the N$\times$N matrix $M$ is time and position
independent.  This implies that $ S^{(0)}(x,x)$ has the same property.
Indeed
\begin{displaymath}
S^{(0)} (x,y) = - \int \frac{d^4p}{(2 \pi)^4} S^{(0)} (p) e^{i p \cdot
(x-y)},
\end{displaymath}
with
\begin{displaymath}
S^{(0)} (p)= \frac{i}
{(p_{\mu} + q_{\mu} \tau_y)(p^{\mu} + q^{\mu} \tau_y)
-M^2 + i \varepsilon},
\end{displaymath}
as can be checked by comparing the action of the two operators on a plane
wave state $\exp ( ik \cdot y )$.

For the particular form we assumed for ${\bar \varphi}$ and $S(x,y)$ the
self consistent mass reads
\begin{eqnarray*}
m^2 (x) & = & U(x) \left[ m_0^2 +\frac{\lambda}{6} \varphi_0^2 +
\frac{\lambda}{6} {\rm tr} S^{(0)} (x,x) \right. \\
& & \qquad + \left. \frac{\lambda}{3} {\bar \varphi}^{(0)} \times {\bar
\varphi}^{(0)} + \frac{\lambda}{3} S^{(0)} (x,x) \right]
U^{\dagger}(x) \mbox{.}
\end{eqnarray*}
Using the formula
\begin{displaymath}
U(x) \partial_{\mu} U^{\dagger}(x) = \partial_{\mu} - i q_{\mu} \tau_y,
\end{displaymath}
it can be seen that the previous equation implies that our Ansatz
actually solves the mean field equations provided the mass matrix $M$
in the propagator $S^{(0)}$ satisfies
\begin{eqnarray*}
M^2 &=& m_0^2 +\frac{\lambda}{6} \varphi_0^2 
+ \frac{\lambda}{6} {\rm tr} S^{(0)} (x,x) 
\\ 
& & \qquad 
+ \frac{\lambda}{3} {\bar \varphi}^{(0)}\times {\bar \varphi}^{(0)} 
+ \frac{\lambda}{3} S^{(0)} (x,x) \mbox{.}
\end{eqnarray*}
This N$\times$N matrix equation defines the mass gap $M$. It will be
referred to below as gap equation in the rotating frame. To have a
closed set, the previous equations must be supplemented by the
relation satisfied by the condensate ${\bar \varphi}^{(0)}$ in the
rotating frame
\begin{eqnarray*}
& & \left[ -q^2 + m_0^2 + \frac{\lambda}{6} \varphi_0 ^2 +
\frac{\lambda}{6} {\rm tr} S^{(0)} (x,x) \right. \\
& & \left. \qquad \qquad \qquad \qquad 
+ \frac{\lambda}{3} S^{(0)}_{11}(x,x) \right] \varphi_0 =0 \mbox{.} 
\end{eqnarray*}
where we have use the assumption that the condensate in the rotating
frame is position and time independent and takes the form
(\ref{PHI0}).  This equation has a trivial solution $\varphi_0 = 0$
having O(N) symmetry when $q = 0$. The interesting solution however
(at least when $q = 0$) is expected to have a non-vanishing value of
the condensate.

\section{Gap Equation in the Rotating Frame}
\label{sect_gap}

Let us first show that the solution of the gap equation is a diagonal matrix
which is furthermore a multiple of the unit matrix in the subspace
$a$=3,\ldots N i.e. $M^2_{ab}$ = $M_a^2$ $\delta_{ab}$ with $M_a$ = $\mu$ for
$a$=3,\ldots N.

Let us consider the propagator in the subspace of the first two
flavors $S^{(0)} (p)$ in the momentum representation. It can be
written as the following 2$\times$2 matrix
\begin{displaymath}
S^{(0)} (p)= \frac{1}{\Delta} \pmatrix{ p^2+ q^2- M^2_2 & -2ip \cdot q \cr
2i p \cdot q & p^2+ q^2- M^2_1 \cr },
\end{displaymath}
where $\Delta$ is the determinant
\begin{displaymath}
\Delta=  \left( p^2+ q^2- M_1^2 \right) \left(  p^2+ q^2- M_2^2
\right) - 4(p \cdot q)^2.
\end{displaymath}
In order to build the quantity $S^{(0)}(x,x)$ we have to integrate
$S^{(0)}(p)$ over $p$. Since off-diagonal elements are odd functions
of momentum, we see that the assumption of a diagonal mass matrix is
compatible with the structure of the gap equation.

The equations determining the four quantities characterizing the solution
$M_1^2$, $M_2^2$, $\mu^2$ and $\varphi_0$ are the following
\begin{equation} \label{GAP} \begin{array}{lll} \displaystyle
M_a^2 &=& m_0^2 +\frac{\lambda}{6} \varphi_0^2 + 
\frac{\lambda}{6} {\rm tr} S^{(0)}(x,x) \\
&& \\
&+& \frac{\lambda}{3} 
\left[ S_{aa}^{(0)}(x,x) + \delta_{a,1} \varphi_0^2 \right] \mbox{,} 
\end{array}
\end{equation}
while the condensate satisfies
\begin{equation} \label{CONDENSATE}
\left( M_1^2 -\frac{\lambda}{3} \varphi_0^2 -q^2 \right) \varphi_0=0.
\end{equation}
The diagonal elements of $S^{(0)}(x,x)$ have the following expression
\begin{displaymath}
S_{aa}^{(0)}(x,x)=- \int \frac{d^4p}{(2 \pi)^4} S_{aa}^{(0)}(p),
\end{displaymath}
where for the first two flavors $a$=1,2 one has
\begin{equation}\label{S0}
S_{aa}^{(0)}(p)= \frac{i ( p^2+ q^2- M_1^2 -M_2^2 + M_a^2)}
{(p^2+ q^2- M_1^2)(  p^2+ q^2- M_2^2) - 4(p \cdot q)^2 }, 
\end{equation}
where the masses appearing in this equation are supposed to carry a
vanishingly small negative imaginary part.
For the flavors $a$=3,\ldots N one has
\begin{displaymath}
S_{aa}^{(0)}(p)= \frac{i}
{p^2- \mu^2 + i \varepsilon},
\end{displaymath}
a quantity which is manifestly flavor independent. 

These formulae complete the construction of the rotating solutions of the
mean field equations for the sigma model.

\section{Structure of the Gap Equation}
\label{sect_str}

The previous equations involve divergent integrals which need to be
regularized e.g. by the introduction of a cutoff in momentum
$\Lambda$. For $q$=0 one recovers the usual static mean field
equations for the vacuum state.  All cutoff dependence may, in
principle, be removed by the proper renormalization of the mass and
coupling constants.  In the following analysis, we instead treat the
$O(N)$ model as an effective theory retaining the cutoff dependence
explicitly.

In examining the structure of the coupled integral equations
(\ref{GAP}) -- (\ref{S0}) in the presence of non-vanishing $q_{\mu}$,
it is instructive to absorb $q^2$ into the masses by rewriting the
effective masses of the first two flavors by ${\cal M}^2_1 = M^2_1-
q^2$ and ${\cal M}^2_2 = M^2_2 - q^2$, while keeping other masses
unchanged, ${\cal M}_a^2 = M_a^2$ for $a \geq 3$. Then we find, for
the first two flavors,
\begin{eqnarray*}
{\cal M}^2_1 & = & m_0^2 - q^2 +\frac{\lambda}{2} \varphi_0^2 
+ \frac{\lambda}{6} {\rm tr} S^{(0)} (x,x) + \frac{\lambda}{3}
S_{11}^{(0)} (x,x) \\
{\cal M}^2_2 & = & m_0^2 - q^2 +\frac{\lambda}{6} \varphi_0^2 
+  \frac{\lambda}{6} {\rm tr} S^{(0)} (x,x) + \frac{\lambda}{3}
S_{22}^{(0)} (x,x)  
\end{eqnarray*}
where 
\begin{eqnarray*}
S_{11}^{(0)}(p) & = & \frac{i ( p^2 - {\cal M}_2^2)}
{(p^2 - {\cal M}_1^2)(  p^2 - {\cal M}_2^2 ) - 4(p \cdot q)^2 }, \\
S_{22}^{(0)}(p) & = & \frac{i ( p^2 - {\cal M}_1^2)}
{(p^2 - {\cal M}_1^2)(  p^2 - {\cal M}_2^2 ) - 4(p \cdot q)^2 }, \\
\end{eqnarray*}
while the condensate equation becomes
\begin{displaymath}
\left( {\cal M}_1^2 -\frac{\lambda}{3} \varphi_0^2 \right)
\varphi_0=0.
\end{displaymath} 

One can see in this form that a part of the changes of the gap
equation caused by non-vanishing $q$ is to shift the bare mass
parameter $m_0^2 \to m_0^2 - q^2$.  We expect therefore that time-like
$q_{\mu}$ causes an effect equivalent to increase $|m_0^2 |$, hence
it leads to an increase of the amplitude of the condensate.  This effect
may be compared to that of the centrifugal force in ordinary rotation
which would cause elongation of a non-rigid rotating body.  Note that
for time-like $q_{\mu}$ one can find a frame where the spatial
component of $q_{\mu}$ vanishes and the condensate rotates uniformly
with angular frequency $\omega = \sqrt{q^2}$.  On the other hand, for
space-like $q_{\mu}$ there is a frame where the condensate becomes
static and oscillates spatially with the wavelength $\lambda = 2\pi/
\sqrt{-q^2}$.  In this case, the amplitude of the condensate tends to
diminish from its value at $q = 0$.  These effects are purely
kinematical and arises also in the classical limit where the amplitude
of the condensate is determined by
\begin{equation}\label{CLASSICAL}
-q^2 + m_0^2 + \frac{\lambda}{6} \varphi_0^2=0   
\qquad \mbox{ (classical) }. 
\end{equation}

The remaining effect of non-vanishing $q$ is to introduce coupling
between the first two flavor states.  This effect resembles that of
the Coriolis force in nuclear rotation which introduces a coupling
between single particle states with different orbital angular momenta
in a deformed potential.  This is a genuine quantum effect which does
not show up in the classical approximation.

We now proceed to estimate the significance of these effects
by perturbation theory for a small $q^2$.

\section{Perturbative Calculation of $q^2$-dependence}
\label{sect_crit}

Let us denote by $\delta M_a^2$ the change in $M_a^2$ when going from a
square momentum zero to $q^2$. The corresponding change in the diagonal
matrix element of $S^{(0)}$ may be expressed as
\begin{equation}\label{DELTAS} 
\delta  S_{aa}^{(0)}(x,x) = J_a \delta M_a^2 +  K_a q^2 
\end{equation}
The coefficients $J_a$ and $K_a$ can be obtained by using the series
expansion of the propagator (\ref{S0}) and integrating out over the
momentum $p$.  The result is given by
\begin{eqnarray*}
J_1  & = & G' ( M_0^2 ), ~ J_2 = J_3 = \cdots = J_N = G' ( \mu^2 ), \\
K_1  & = & I ( M_0^2, \mu^2 ), ~ K_2 = I ( \mu^2, M_0^2 ), \\
K_3  & = & \cdots = K_N = 0.
\end{eqnarray*}
where we have introduced the following integrals
\begin{eqnarray*}
G (m^2) & = & - \int \frac{d^4p}{(2 \pi)^4} \frac{i} {p^2- m^2 + i
\varepsilon} \\
I (M_a^2, M_b^2) & = & -  M_b^2 \int \frac{d^4p}{(2 \pi)^4}
\frac{i } {(p^2- M_a^2)^2(p^2- M_b^2)} .
\end{eqnarray*}
and defined the function $G'(m^2)$ as the derivative of $G (m^2)$ with
respect to $m^2$.  Here $M_0$ stands for the value of $M_1$ when $q=0$
while the unperturbed masses of all other flavors ($a= 2,\ldots, N$)
are set to $\mu$.  They satisfy the gap equations at $q = 0$. 

The integral $G(m^2)$ is divergent and requires regularization.  Using
a 3-dimensional regularization in momentum space with a cutoff
$\Lambda$ it is found to be
\begin{displaymath}
G(m^2)= \frac{1}{8 \pi^2} 
\left[ \Lambda^2 - m^2 \log \left( \frac{2 \Lambda}{m} \right) + \frac{m^2}{2}
\right] .
\end{displaymath}
Differentiating with respect to the square mass gives
\begin{displaymath}
G'(m^2)= -\frac{1}{8 \pi^2} 
\left[ \log \left( \frac{2 \Lambda}{m} \right) - 1 \right] .
\end{displaymath}
On the other hand, the integral $ I (M_a^2, M_b^2)$ is finite. 
Using the standard integral representation of propagators \cite{ZUBER}, 
the following closed expression for this integral can be obtained
\begin{eqnarray*}
I (M_a^2, M_b^2) &= &\frac{1}{16 \pi^2} \frac{M_b^2}{M_a^2-M_b^2} \\
&& \qquad \times \left[ 1 -
\frac{M_b^2}{M_a^2-M_b^2} \log \left( \frac{M_a^2}{M_b^2} \right) \right],
\end{eqnarray*}

The linearization of the gap equation leads to the following relations
for the mass changes in the first two flavors
\begin{displaymath}
\delta M_1^2= \frac{\lambda}{2} \delta \varphi_0^2 +
\frac{\lambda}{6}  {\rm tr} \delta S^{(0)} (x,x) +
\frac{\lambda}{3} \left[ J_1 \delta M_1^2  + K_1 q^2 \right] ,
\end{displaymath}
and
\begin{displaymath}
\delta M_2^2= \frac{\lambda}{6} \delta \varphi_0^2 +
\frac{\lambda}{6}  {\rm tr} \delta S^{(0)} (x,x)+
\frac{\lambda}{3} \left[  J_2 \delta M_2^2 + K_2 q^2 \right] .
\end{displaymath}
For other flavors one finds
\begin{displaymath}
\delta \mu^2= \frac{\lambda}{6} \delta \varphi_0^2 +
\frac{\lambda}{6} {\rm tr} \delta S^{(0)} (x,x)+
\frac{\lambda}{3} J_2 \delta \mu^2.
\end{displaymath}
A closed system of linear equations can be constructed 
by including the linearized form of the evolution equation for the
condensate (\ref{CONDENSATE})
\begin{equation} \label{LCONDENSATE}
\delta M_1^2 - \frac{\lambda}{3} \delta \varphi_0^2 -q^2=0,
\end{equation}
and the expression for the total change in the trace of the propagator
\begin{eqnarray*}
{\rm tr} \delta S^{(0)} (x,x) &=& J_1 \delta M_1^2 
+ J_2 \left[ \delta M_2^2 + (N - 2) \delta \mu^2 \right] \\ 
& & \qquad +  (K_1+K_2)q^2.
\end{eqnarray*}
The solution of this linear set gives for the changes in the masses
\begin{displaymath}
\delta M_1^2= (c_1 b_2-b_1 c_2) \frac{q^2}{\Delta}, 
\end{displaymath}
and
\begin{displaymath}
\delta \mu^2= (a_1 c_2-c_1 a_2) \frac{q^2}{\Delta}. 
\end{displaymath}
The constants appearing in these equations are
\begin{eqnarray*}
a_1 & = & \frac{\lambda}{3} J_1 ,~~~a_2= 1 + \frac{\lambda}{3} J_1 , \\
b_1 & = & 1 - \frac{\lambda}{3} J_2 ,~~~b_2=  (N+1) \frac{\lambda}{3}
J_2 - 2, \\
c_1 & = & 1 - \frac{\lambda}{3} K_1, ~~~ 
c_2  =  1  - \frac{\lambda}{3} K_1 - \frac{\lambda}{3} \frac{K_2}{b_1} .
\end{eqnarray*}
The quantity $\Delta$ in the above equations is the determinant of the linear
system
\begin{displaymath}
\Delta=  (N+2) \left( \frac{\lambda}{3} \right)^2 J_1 J_2 - \lambda
J_1 + \frac{\lambda}{3} J_2 - 1. 
\end{displaymath}
The change in the condensate is given by equation (\ref{LCONDENSATE}).

\section{Discussion}
\label{sect_dis}

In the above result all quantum effects are contained in the
parameters $J_a$ and $K_a$ which represent the changes in the vacuum
fluctuation caused by non-vanishing $q$.  The classical relation
(\ref{CLASSICAL}) can be obtained from this result by setting $J_a =
K_a = 0$.  In this limit one finds 
\begin{equation} \label{CCONDENSATE}
\frac{\lambda}{3} \delta \varphi_0^2 = 2 q^2 \qquad \mbox{ (classical) 
}.
\end{equation}

To obtain an estimate of the quantum fluctuations we have chosen $M_0$
to be of the order of the sigma mass $M_{\sigma}$ i.e. about 500 MeV
and set $N=4$. Assuming the value of the condensate $\varphi_0$ to be
the pion decay constant $f_{\pi} = 93$ MeV, the condensate equation at
$q=0$ gives a coupling constant $\lambda = 86.7$.  For a momentum
cutoff $\Lambda$ = 1 GeV, the gapequation gives $\mu = $ 224 MeV.
Note that $\mu$ is a variational parameter characterizing the
significance of the quantum fluctuation of pion fields; it should not
be taken as the pion mass.  For the previous values of the masses one
has $J_1$=-.00489, $J_2$=-.0141, $K_1$=.000948 and $K_2$= 0.00800.
Although the coupling constant is large compared to unity the
constants $K_1$ and $K_2$ are sufficiently small to have $c_1$ and
$c_2$ very close to unity. Actual values are
\begin{displaymath}
c_1=.973,~~~~~c_2=.812,
\end{displaymath}
which means that the most important {\it explicit} $q^2$ dependence is
the one arising in the equation for the condensate. The above values
give the following expressions for the variations of the masses with
momentum
\begin{displaymath}
\delta M_1^2= 8.12 q^2 ~~~{\rm and}~~~
\delta \mu^2= 1.48 q^2.
\end{displaymath}
The change in the condensate is found to satisfy 
\begin{equation} \label{VCONDENSATE} 
\frac{\lambda}{3} \delta \varphi_0^2 = 7.15 q^2.
\end{equation} 
An increase of the angular velocity (time-like $q^2$) thus produces an
increase of the chiral radius, both in the classical approximation and
in the mean-field picture, in agreement with the qualitative argument
given above. The change is more rapid in the quantum case. Similarly
for a space-like $q^2$ a vanishing value of the condensate is reached
more rapidly in the quantum case.  This is expected because quantum
fluctuations smear out the effective potential and make symmetry
breaking more difficult to reach.  In the quantum case a transition
occurs when
\begin{equation} \label{CRITICALQ} 
- q^2 = q_c^2 \simeq M_{\sigma}^2/7. 
\end{equation}
At this point the mass of the first flavor vanishes which implies that
a solution with a non zero condensate can no longer be obtained beyond
this point.  The corresponding classical excitation energy density of
the meson field is $-q_c^2$ $f_{\pi}^2$/2 $\simeq$ (123 MeV)$^4$.  

It is interesting to compare this critical momentum for space-like
condensate to the corresponding value in the classical limit obtained
from (\ref{CLASSICAL}) for $\varphi_0 = 0$:
\begin{displaymath}
q_c^2 = M_{\sigma}^2/2 \qquad \mbox{ (classical) } 
\end{displaymath}
where we have used the relation $M_{\sigma}^2 = - 2 m_0^2$ in the tree
(classical) approximation.  The coupling to the quantum fluctuation
thus works to suppress the appearance of static condensate with longer
wavelengths.  This implies that the appearance of the static chiral
condensate is more sensitive to the size of the system than the
classical treatment predicts.  This would have important implications
for the dynamics of chiral condensate in high energy nuclear 
collisions. 

It is worthwhile noting that the analogy with nuclear deformation is
limited by the fact that in the nuclear case deformation disappears in
the classical limit contrary to symmetry breaking in the $O(N)$ model.
Evolutions with angular velocity are however similar.

In conclusion we have shown that, as compared to analyses based on
classical equations, new effects arise in a quantum framework as a
result of the coupling between the evolutions of mean values and
fluctuations.  Our result indicates that the coupling to the quantum
fluctuation suppresses static condensate with spatial oscillation
while the condensate oscillating in time is amplified by the quantum
effects.

{\bf Aknowledgements}

One of us (D. V.) wishes to express his appreciation to the Japan
Society for the Promotion of Science for support to visit the Yukawa
Institute where this work was started.  The other (T. M.) is grateful
to Institute de Physique Nucl\'eaire at Orsay for the support and the
hospitality which he received during his visit.

\end{document}